\documentclass[sii]{ipart}

\usepackage{graphicx}
\usepackage{subfigure}
\usepackage{colortbl,color,array} % used here to color a column
\usepackage{ctable,multirow}

\usepackage{amsthm,amsmath}
\usepackage[numbers,square]{natbib}
\RequirePackage[dvips]{hyperref}
\usepackage{color}

\def\min{\mathop{\rm min}}

\def\mean{\mathop{\rm mean}}

\RequirePackage{hypernat}

\startlocaldefs
\endlocaldefs

\begin{document}

\begin{frontmatter}

% "Title of the Paper"
\title{Analyzing LC-MS/MS data by spectral count and ion abundance: two case studies}
%\thanksref{t1}}
\runtitle{Analyzing by spectral count and ion abundance}

\begin{aug}
\author{\fnms{Thomas I.} \snm{Milac}\thanksref{t2}\corref{}\ead[label=e1]{milac@uw.edu}},
\address{Department of Pharmacology \\
Box 357280 \\
University of Washington \\
Seattle, WA 98195-7280 \\
\printead{e1}}
\author{\fnms{Timothy W.} \snm{Randolph}\ead[label=e2]{trandolph@fhcrc.org}}
\address{Fred Hutchinson Cancer Research Center \\
1100 Fairview Ave. N. \\
PO Box 19024 \\
Seattle, WA 98109 \\
\printead{e2}}
\and
\author{\fnms{Pei} \snm{Wang}\ead[label=e3]{pwang@fhcrc.org}}
\address{Fred Hutchinson Cancer Research Center \\
1100 Fairview Ave. N. \\
PO Box 19024 \\
Seattle, WA 98109 \\
\printead{e3}}

\thankstext{t2}{to whom correspondence should be addressed}

\runauthor{Milac et~al.}
\end{aug}

% history:
%\received{\smonth{1} \syear{0000}}

\begin{abstract}
In comparative proteomics studies, LC-MS/MS data is generally quantified using one or both of two
measures: the spectral count, derived from the identification of MS/MS spectra, or some measure of ion
abundance derived from the LC-MS data.  Here we contrast the performance of these measures and
show that ion abundance is the more sensitive.  We also examine how the conclusions of
a comparative analysis are influenced by the manner in which the LC-MS/MS data is `rolled up' to the
protein level, and show that divergent conclusions obtained using different rollups can be informative.
Our analysis is based on two publicly available reference data sets, BIATECH-54 and CPTAC, which
were developed for the purpose of assessing methods used in label-free differential proteomic studies.
We find that the use of the ion abundance measure reveals properties of both data sets not readily
apparent using the spectral count.
\end{abstract}

\begin{keyword}[class=AMS]
\kwd[Primary ]{62P10}
%\kwd{}
\kwd[; secondary ]{92D20}
\end{keyword}

\begin{keyword}
\kwd{mass spectrometry}
\kwd{comparative proteomics}
\kwd{ion abundance}
\kwd{spectral count}
\kwd{ion competition}
\end{keyword}

%\tableofcontents

\end{frontmatter}

\section{Introduction}
Comparative proteomics studies aim to discern differences in protein content and abundance
between case and control samples.  Tandem liquid-chromatography mass-spectrometry (LC-MS/MS)
experiments are performed routinely in carrying out such studies and may employ labeled or
unlabeled samples.  Our focus here will be on the analysis of data from unlabeled experiments.

Preparatory to an LC-MS/MS experiment, a protein sample is digested using trypsin or other
proteolytic enzyme.  In LC-MS, a reverse phase liquid chromatography (LC) column is typically
used to separate the resulting peptide `species' based on their hydrophobicity, and
an MS spectrum, or scan, is taken of them periodically as they elute. The collection of scans from a
single experiment may be viewed as a three-dimensional landscape of peaks located in
elution time ($t$) and mass/charge ($m/z$) space.  We refer to the recent reviews \cite{Deutsch:2008}
and \cite{Domon:2010} for a more detailed discussion of LC-MS/MS experimental and analytical
procedures.

The peaks associated with a single species form a characteristic group, or `feature', in LC-MS
space.   An example of such a feature is shown in Figure~\ref{fig:MS1_feature}. Each peak in
the group collects ions of one or more isotopic forms of the species. The relative amplitude of
the peaks measures the relative abundance of the isotopic forms.  A number of software packages
have been developed to detect and quantify LC-MS features including MapQuant \cite{Leptos:2006},
MaxQuant \cite{Cox:2008}, Sahale \cite{Milac:2012}, Serac \cite{Old:2005}, SpecArray \cite{Li:2005},
and SuperHirn \cite{Mueller:2007}.

In tandem MS (MS/MS), selected LC-MS peaks are interrogated by a collision-induced dissociation
(CID) and a second MS scan recorded.  Figure~\ref{fig:MS1_feature} displays the $(t,m/z)$ locations
(shown in red) of four MS/MS scans sampled from the LC-MS feature. MS/MS spectra are matched against
a protein database to determine the species from which they likely originated. In the simplest case, the
species is one of the amino acid sequences generated by an \emph{in silico} digest of the protein
database, e.g., by trypsin.  The search space, however, is usually
expanded beyond that of the \emph{in silico} digest to include, for example, species with missed or
non-tryptic cleavages, and species with altered mass due to anticipated chemical modifications of specific
amino acids.  Algorithms to perform this search are implemented by software packages including
MyriMatch \cite{Tabb:2007}, Sequest \cite{Eng:1994}, and X!Tandem \cite{Craig:2004}.  Following
identification, each spectrum-to-species match is assigned an instrument-independent quality score,
typically either a PeptideProphet score \cite{Keller:2002}, or a false discovery rate (FDR) calculated
on the basis of hits against decoy proteins in the database searched \cite{Kall:2008}.

In the analysis of LC-MS/MS data, the relative abundance of species in a sample is generally quantified
by either or both of two measures: the spectral count (see, e.g, \cite{Choi:2008,Liu:2004,Lu:2006,Zhang:2006}),
which is the number of MS/MS spectra identified as arising from the species, or some measure of the
species' ion abundance derived from an analysis of its feature signature in LC-MS space (see, e.g.,
\cite{Bondarenko:2002,Jaffe:2006,Leptos:2006,Old:2005}). For example, the spectral count
associated with the species giving rise to the feature in Figure~\ref{fig:MS1_feature} is four, and one
possible measure of the species' ion abundance is the volume of the peaks fitted to its feature, as shown
in Figure~\ref{fig:Sahale_fit}.  Recently, the ion abundance of MS/MS spectra has been introduced as one
component of an alternative quantitative measure \cite{Griffin:2009}; we do not consider this measure of
abundance in this work.

We note that although it is the species that is observed directly in a LC-MS/MS experiment, and for which
such experiments yield quantitative data, the goal of comparative proteomics studies is to identify
and quantify the proteins. In particular, the data available for species must be `rolled up'
to the protein level.  To date, no systematic research has addressed how best to infer protein quantity
from the species quantities \cite{Podwojski:2010}, although some statistical discussion on this topic is
provided by \cite{Clough:2009} for ion abundance and \cite{Lundgren:2010} for spectral count.  A common
approach is to average the quantitative measure used (i.e., spectral count or ion abundance) for all species
belonging to a particular protein and then to use the result as a surrogate measure for the abundance of
the protein in a statistical test. Other investigations infer relative protein abundance directly from the
abundance of species, or from the abundance of species data rolled up to some intermediate level,
using a variety of methods. For methods based on spectral counts, see APEX \cite{Lu:2006},
emPAI \cite{Ishihama:2005}, QSpec \cite{Choi:2008}, SASPECT \cite{Whiteaker:2007,WangSASPECT:2008}
and Spectral Index \cite{Fu:2008}.  For statistical models of protein rollup based on ion abundances,
see \cite{Clough:2009, Karpievitch:2009}. The recent review \cite{Neilson:2011}
provides additional perspective.

In this paper we contrast the performance of the spectral count and ion abundance in quantifying
LC-MS/MS data.  We also consider the effect of various levels of data aggregation---from the species
level to the protein level---prior to, or simultaneously with, the analysis of the relative abundance of
proteins. We  conclude that ion abundance, coupled with an appropriate rollup procedure, is the more
sensitive measure for use in comparative analysis.  Our findings are based on detailed examinations
of two publicly available reference data sets,  BIATECH-54 \cite{Kolker:2007} and
CPTAC \cite{Paulovich:2010}, which were developed to assess methods of protein identification and
quantification in LC-MS/MS experiments. The use of the ion abundance measure reveals characteristics
of both the BIATECH-54 and CPTAC data sets not readily apparent by the use of the spectral count.

\section{Methods}
\subsection{Quantification}
\label{subsection:Quantification}
\begin{figure}[!h]
 \begin{center}
  \subfigure[]
      {\label{fig:MS1_feature}
        \includegraphics[scale=0.9]{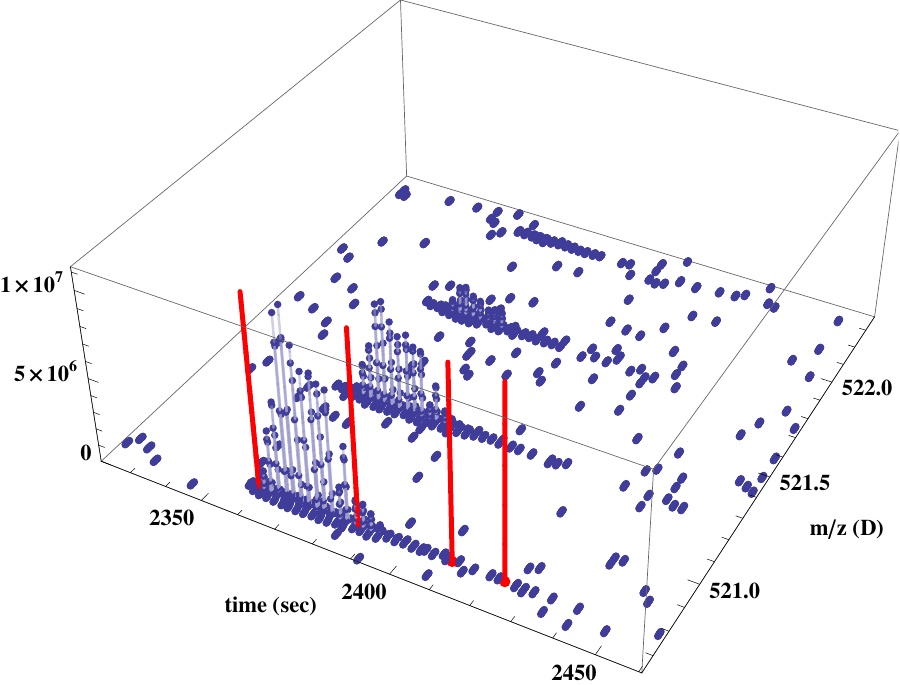}} \\
  \subfigure[]
      {\label{fig:Sahale_fit}
        \includegraphics[scale=0.9]{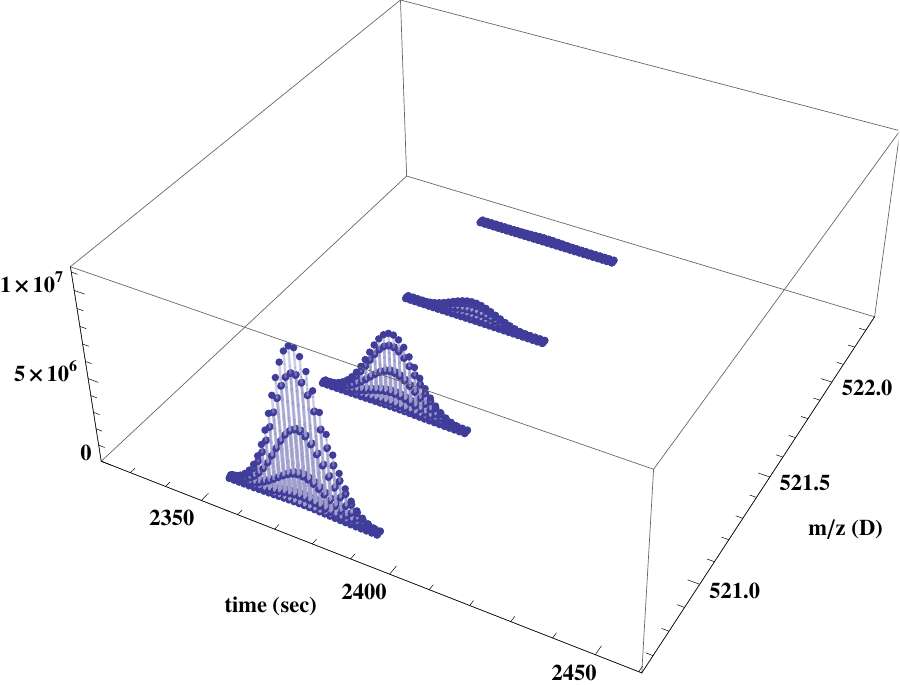}}
 \end{center}
 \vspace{-0.1in}
 \caption{{\bf An LC-MS feature and its quantification.} (a) The LC-MS feature associated with the species
 GGALDFADFK+2 identified in file 54ProtMix\_1ug\_051207\_SampA1\_01.mzXML of the BIATECH-54 data set
 \cite{Kolker:2007}.  Red lines indicate the location at which MS/MS spectra were taken, and later identified
 as originating with this species by X!Tandem. (b) The fit to this feature made by Sahale \cite{Milac:2012}.}
 \label{fig:LC-MS}
\end{figure}

We used Sahale \cite{Milac:2012} to determine the spectral count and ion abundance
of species identified by X!Tandem in the BIATECH-54 and CPTAC data.  Briefly, Sahale searches for
LC-MS features in the vicinity of the $(t, m/z)$ location of those species identified by MS/MS satisfying
a specified quality threshold, either a PeptideProphet score or false discovery rate (FDR). The species'
spectral count is the number of MS/MS identifications of that species satisfying the
threshold.  The species' ion abundance is determined by fitting the corresponding feature to
the model $f \equiv f(t, m)$,
\begin{equation}
 f = A \overbrace{\exp \left[ -\frac{(t - \mu)^2}{2 \sigma^2} \right]}^{\mathrm{model \; in \;} t}
                 \overbrace{ \left( \sum_{k=0}^{N-1} \frac{\lambda^k}{k!} e^{-\lambda}
                                        \exp\left[ -\frac{(m-\zeta_k)^2}{2 \rho^2} \right] \right)}^
                                                          {\mathrm{model \; in \;} m/z},
 \label{eq:Feature function}
\end{equation}
and taking the total volume under the fitted surface,
\begin{equation*}
 2 \pi A  \sigma \rho \sum_{k=0}^{N-1} \frac{\lambda^k}{k!} e^{-\lambda},
 \label{eq:Ion abundance}
\end{equation*}
as a surrogate measure of ion abundance.
$f(t,m)$ is the product of a simple Gaussian in the chromatographic ($t$) coordinate and a
series of $N$ Gaussians with Poisson-distributed peak amplitudes in the $m/z$ ($m$) coordinate.
In Equation~\eqref{eq:Feature function}, $A$ is the amplitude.  $\mu$ is the coordinate of the
peak of the time Gaussian and $\sigma$ is its standard deviation. For the function in $m/z$,
$N$ is the number of isotopic peaks modeled, $\zeta_k = \zeta_0 + k \delta$ is the $m/z$ location of the
$k$th peak, $\zeta_0$ is the coordinate of the first Poisson-distributed peak, $k$ is the peak
number, $\delta$ is the inter-peak spacing, $\lambda$ is the Poisson parameter, and $\rho$ is the
standard deviation of the Gaussians.  Sahale fits Equation~\eqref{eq:Feature function} to data by least
squares using the Levenberg-Marquardt minimization algorithm. Figure~\ref{fig:Sahale_fit} shows
the fit determined by Sahale to the feature shown in Figure~\ref{fig:MS1_feature}.

Sahale returns the spectral count and ion abundance for every species for which at least one
MS/MS identification exists satisfying the quality threshold specified.  Because Sahale quantifies LC-MS
features guided by MS/MS identifications, and not all the corresponding ion abundance quantifications
are determined successfully, the number of species quantified by ion abundance is less than or
equal to the number quantified by spectral count.

\subsection{Rollup and significance analysis}
\label{subsection:Rollup}
The data collected in an LC-MS/MS experiment quantifies the species.  To determine the
relative abundance of proteins, a statistical analysis might use the quantitative data for species
directly, or will first aggregate the species-level quantitative data to some higher rollup level.

A species as we've defined the term is the identity associated with a feature in LC-MS space
by a search of a protein database, e.g., by X!Tandem.  A species is characterized by three attributes:
its primary amino acid sequence, any chemical modifications to its amino acids, and its charge state.
The nomenclature DEDTQAM[147.035]PFR+2, for example, identifies a species with the primary
amino acid sequence DEDTQAMPFR containing an oxidized methionine at position 7 with a net mass
of 147.035 Daltons (D) and carrying a charge of +2 induced by the LC-MS instrument.  A like species with
charge +3, DEDTQAM[147.035]PFR+3, might be identified in the same experiment.  Other species might
be identified in the experiment with the same primary amino acid sequence but without the modification
to methionine, i.e., DEDTQAMPFR+2 and DEDTQAMPFR+3.  In general, these species variants, all
of which have the same primary amino acid sequence, will correspond to widely-separated features
in LC-MS space.

The key point is that as a result of biological processes and the action of the LC-MS instrument,
a short segment of a protein's primary amino acid sequence may present itself as any number of
species - and corresponding features - in an LC-MS/MS experiment.

We define three levels of rollup of LC-MS/MS data.
\begin{itemize}
 \item \emph{Species}: The base rollup level.
 \item \emph{Peptide}: The collection of all species with the same primary amino acid sequence. The
 peptide DEDTQAMPFR, for example, refers to the species DEDTQAMPFR+2, DEDTQAMPFR+3,
 DEDTQAM[147.035]PFR+2, etc.
 \item \emph{Protein}: The collection of all species that by virtue of their primary amino acid sequence
 possibly originated from a particular protein.
\end{itemize}
Where it is convenient, we refer to the rollup level by a single-letter abbreviation: species (s),
peptide (p), or protein (P).  Note that we use the term `protein' to refer to both a rollup level
and to the biological entity; the meaning will be clear from the context.

We assigned to the higher rollup entities - the peptides and proteins - a spectral count and ion abundance
by a simple sum. For example, we computed the spectral count (ion abundance) for a particular
peptide to be the sum of the spectral counts (ion abundances) for all species that are elements of
the peptide.

To test for proteins present in differing abundance, we computed a statistic, $\tau_r$, that
integrates the evidence of difference, $w$, measured for each of a protein's $K$ constituent
elements identified at rollup level $r$,
\begin{equation}
 \tau_r = \tau_r(w_k, k=1,K) = \frac{1}{K} \sum_{k=1}^K w_k.
 \label{eq:tau_definition}
\end{equation}
Note that $w_k$ is $w$ for the $k$'th element of a protein at rollup level $r$.   $\tau_r$
is simply the mean of the $w$'s measured for a protein's $r$-level rollup elements.

For the element-level measure of difference, $w$, we employed a zero-centered and scaled
version of the Wilcoxon test statistic, $W$, which is the sum of ranks of observations
in the case group in a case-control comparison. $W$ naturally falls in the range
\begin{equation*}
 \left[ W_{\mathrm{min}}, W_{\mathrm{max}} \right] = \left[ \frac{n}{2} (n+1), \frac{n}{2} (n+2m+1) \right],
\end{equation*}
where $n$ and $m$ are the number of case and control observations, respectively, being
compared. In our analysis, we took the `observations' to correspond to the biological
replicates.  We define $w$ as
\begin{equation}
 w = \frac{2}{W_{\mathrm{max}} - W_{\mathrm{min}}} \left[ W - \left(\frac{W_{\mathrm{max}} + W_{\mathrm{min}}}{2}\right) \right].
 \label{eq:w_definition}
\end{equation}
Thus defined, $w \in [-1, 1]$.  Henceforth, for brevity, we refer to the scaled and re-centered
Wilcoxon statistic $w$ simply as the Wilcoxon.  As a consequence of the definition of $w$,
$\tau_r$, defined in Equation~\eqref{eq:tau_definition}, falls in the range $[-1, 1]$ independently
of $K$. Note that at the protein level of rollup, $K = 1$ and $\tau_P = w$, where $w$ is the
Wilcoxon computed for the protein at the protein level of rollup.

We computed $\tau_r$ using both the spectral count and ion abundance measures
for all elements at the species, peptide and protein rollup levels.  To derive p-values, we
computed a null distribution for $\tau_r$ by repeatedly permuting the labels of the case
and control samples (1500x in both the BIATECH-54 and CPTAC studies).

Our aim is to contrast quantification by spectral count and ion abundance as transparently as possible.
This demands that the statistical summaries we employ must be comparable and based on the same
concepts. The statistic $\tau_r$ is simply the average of Wilcoxon test statistics for a protein's
constituent elements at a given rollup level, computed in the same way at each level using the
spectral count or the ion abundance.

\section{Case studies}
\subsection{BIATECH-54}
\begin{figure}[!h]
 \begin{center}
   \includegraphics[width=0.9\linewidth]{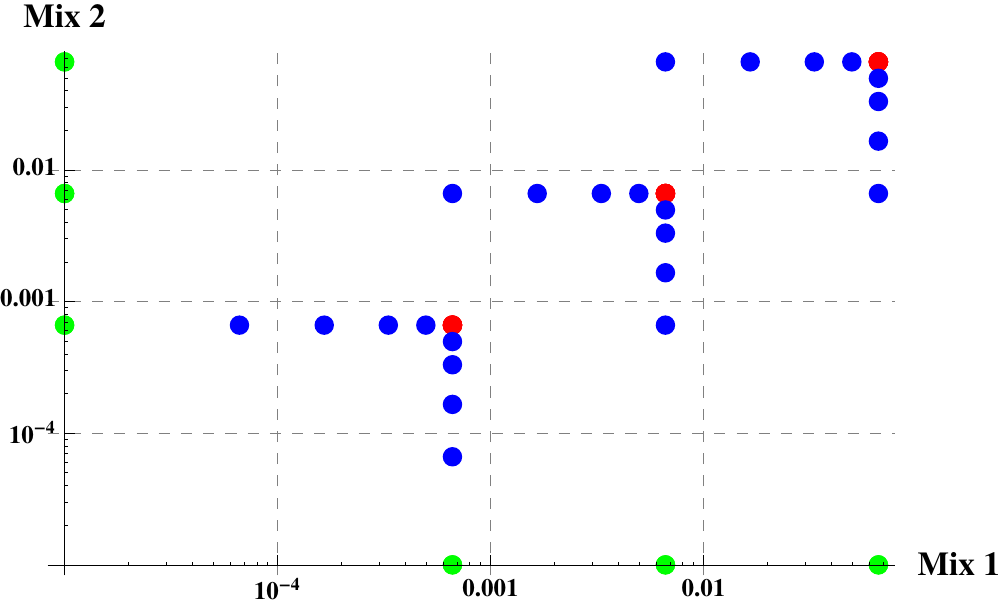}
 \end{center}
 \vspace{-0.1in}
 \caption{{\bf BIATECH-54 study design.} The log milligram (mg) fractional abundances
 (mg of protein/total mg in the sample) of proteins in Mix 1 and Mix 2 of the BIATECH-54 set.
 The \textcolor{blue}{blue}, \textcolor{green}{green} and \textcolor{red}{red} points represent 1, 2
 and 6 proteins, respectively (54 total).  Those proteins for which zero mg was included in either
 sample are shown on the axes.}
 \label{fig:Biatech54_design}
\end{figure}

We describe first our analysis of the LC-MS/MS data set measuring the two reference samples, Mix 1 and
Mix 2, of the BIATECH-54 set (\citet{Kolker:2007}).  The BIATECH-54 mixtures were developed
to assess methods of protein identification and quantification in LC-MS/MS experiments.  Mix 1 and
Mix 2 combine 54 proteins at different concentrations; their exact compositions are documented in
Table 1 of \cite{Kolker:2007} and represented graphically in Figure~\ref{fig:Biatech54_design}. To mimic
biological replicates, Kolker \emph{et al}.\ prepared six (6) replicate samples of Mix1 and Mix 2 and digested
each independently with trypsin.  The twelve (12) samples were then each subject to LC-MS/MS analysis twice;
these are the technical replicates.

The BIATECH-54 data set was obtained from the authors as a collection of 24 files in .mzXML format
\cite{Kolker:2007}. In this collection, files corresponding to the Mix 1 samples are designated
'A'  and Mix 2 samples designated 'B'.

We searched the MS/MS spectra of the data set using X!Tandem \cite{Craig:2004} against a protein
database containing the primary amino acid sequences of the proteins of yeast along with those of
the BIATECH-54 and potential contaminating proteins.  To support estimation of the false discovery
rate, decoy protein sequences, constructed by reversing each of the target protein sequences,
were appended to this database \cite{Elias:2007}.  To compensate for drift in the calibration of the
LTQ-FT instrument used to collect these data, the precursor matching tolerance was set to
an asymmetric window [-10 ppm, +40 ppm].  The search was performed to allow potential oxidation of
methionines ($\Delta$mass = 15.9949D).  Up to two missed tryptic cleavages were permitted.

We conducted separate searches allowing for only fully-tryptic species and for semi-tryptic
species.  As implemented by X!Tandem, the latter are species constrained to have a tryptic terminus
on at least one end.  We chose to analyze the results obtained from both the fully- and semi-tryptic
searches because it is not uncommon in practice to do one or the other.  Though usually not stated
explicitly, the rationale for using a semi-tryptic search is that such species are likely to be
present in samples, e.g., due to the activity of contaminating proteolytic enzymes, and ignoring
such species would leave potentially useful quantitative data `on the table'.

The BIATECH-54 data was quantified by spectral count and ion abundance using Sahale
\cite{Milac:2012}.  The FDR threshold parameter set for Sahale was selected to be 0.001
(see Section~\ref{subsection:Quantification}).  Species not observed or quantified in at
least three (3 of 12) of the Mix 1 samples or three (3 of 12) of the Mix 2 samples were filtered
out, as were uninformative species in the ion abundance data for which the ion abundances are
effectively zero or are missing for all samples. The spectral counts and ion abundances for the
species remaining were then rolled up to the peptide and protein levels (see
Section~\ref{subsection:Rollup}).  The resulting six data sets - the spectral counts and ion
abundances across the three rollup levels - were then normalized by sample to make the data
comparable across samples.  The spectral counts were normalized by the total spectral count, and
ion abundances by the median ion abundance.

As already described, the BIATECH-54 data set includes two technical replicate observations of
each of the 12 biological samples measured.  The technical replicates were `averaged'
to obtain a value for the spectral count and ion abundance for the corresponding biological
replicate.  The average was computed so that if one of the technical observations was found to
be missing, the quantity found for the other was assigned as that for the biological replicate.

\begin{figure}[!h]
 \begin{center}
   \subfigure[]
      {\label{fig:Biatech54_fully_ROC}
       \includegraphics[width=0.9\linewidth]{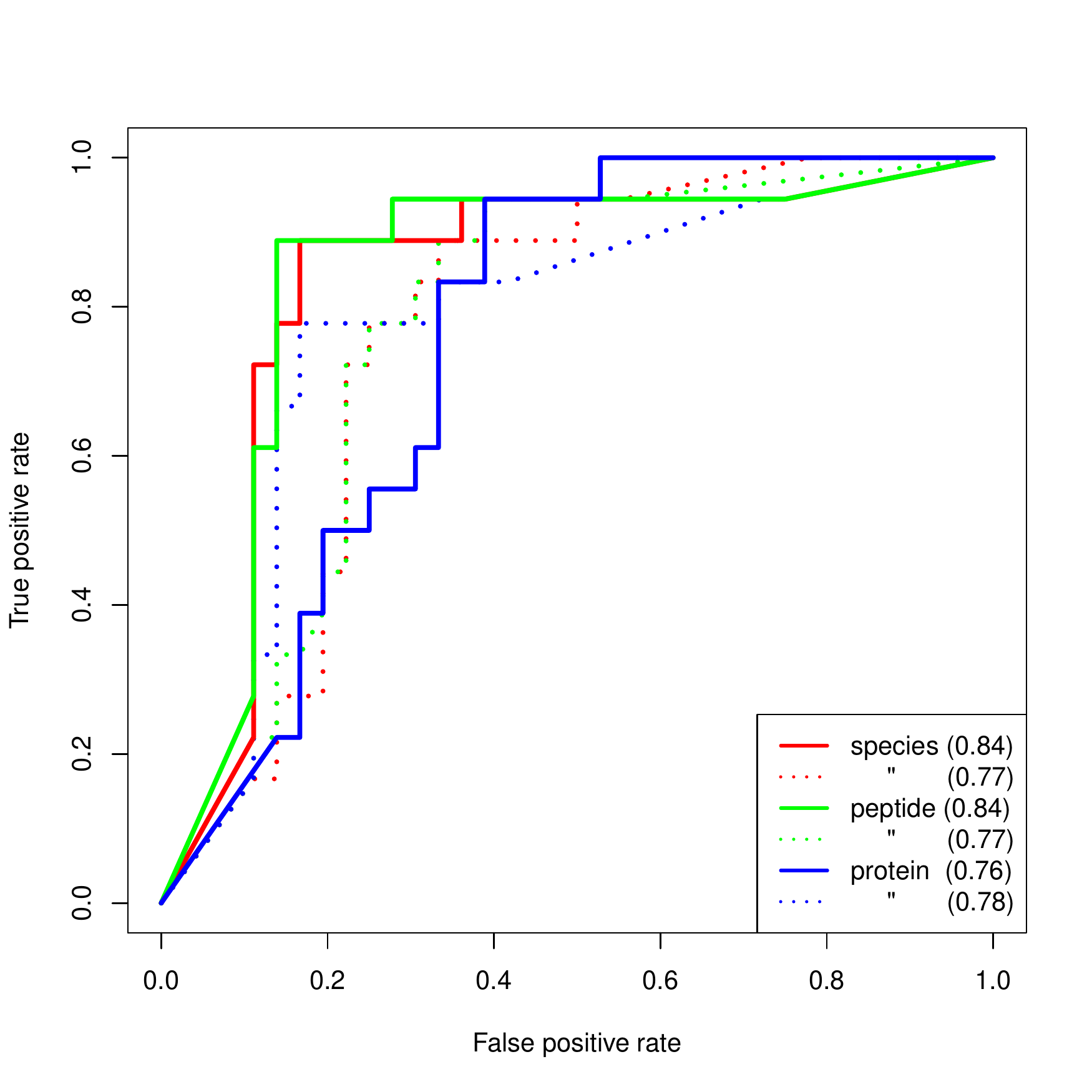}} \\
       \vspace{-0.3in}
   \subfigure[]
      {\label{fig:Biatech54_semi_ROC}
        \includegraphics[width=0.9\linewidth]{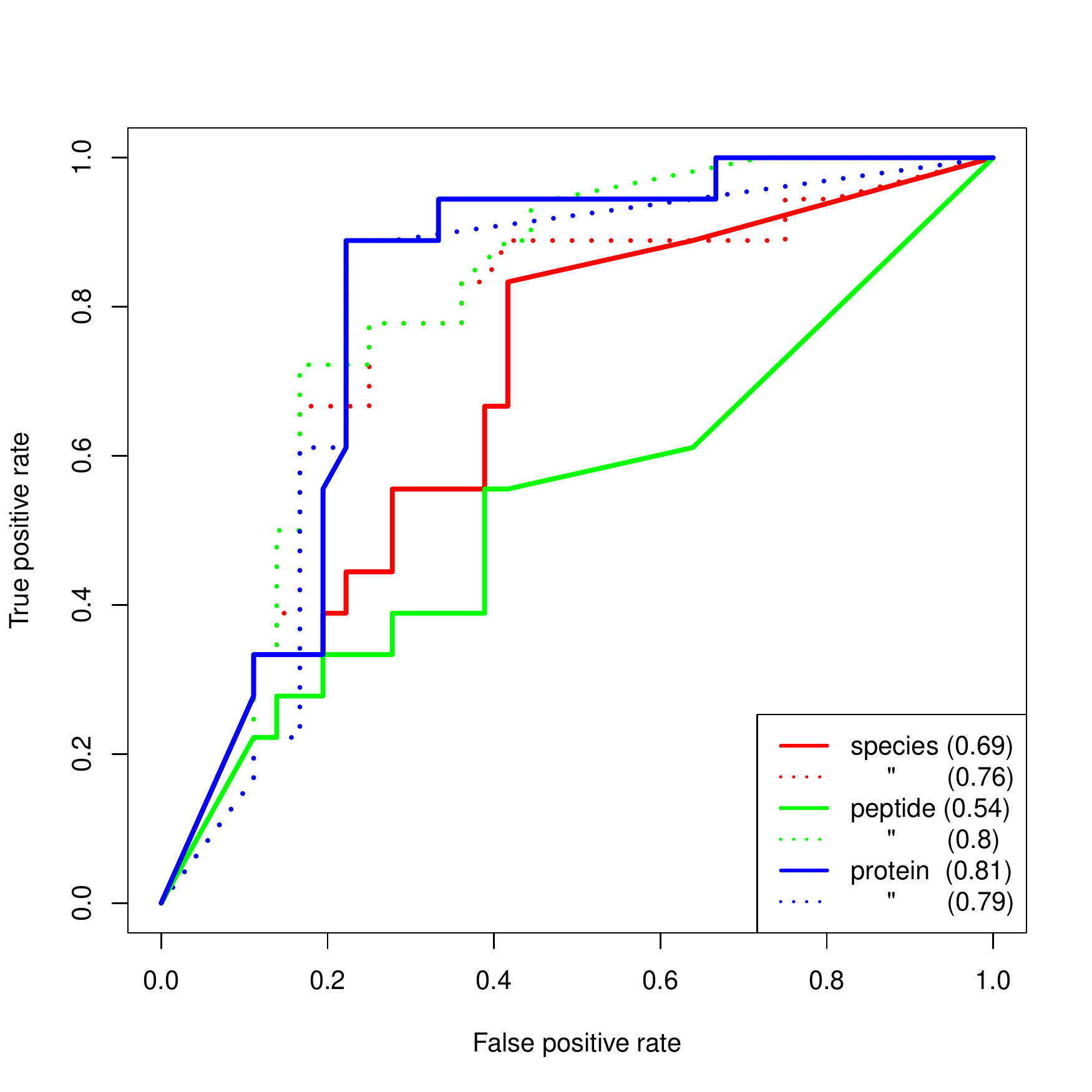}}
 \end{center}
 \vspace{-0.1in}
 \caption{{\bf ROC comparison of performance.} ROC curves showing the performance of $\tau_r$,
 computed using the spectral count (dotted) and ion abundance (solid) measures, in characterizing the
 relative abundance of proteins in Mix 1 and Mix 2 of the BIATECH-54 set across the species, peptide
 and protein rollup levels.  (a) shows the results obtained using the data searched for fully-tryptic species, and
 (b) for semi-tryptic species. The AUC corresponding to each curve is shown in parenthesis.}
 \label{fig:Biatech54_ROC}
\end{figure}

Following these preparatory steps, we tested the species, peptide and protein level entities in
Mix 1 and Mix 2 for evidence of difference using the statistic $\tau_r$
(Equation \eqref{eq:tau_definition}).

Figure~\ref{fig:Biatech54_ROC}, displays ROC curves characterizing the performance
of $\tau_r$ in correctly characterizing the relative abundance of proteins in the BIATECH-54
samples.  Figure~\ref{fig:Biatech54_fully_ROC} shows ROC curves determined by analysis of
the data searched for fully-tryptic species; Figure~\ref{fig:Biatech54_semi_ROC} shows the same
results for data searched allowing for semi-tryptic species. The data used to produce both figures
was quantified by Sahale with the FDR threshold parameter set to 0.001; qualitatively similar results
were obtained with the FDR level set to 0.01 or 0.02 (not shown). The ROC analysis was performed
using the ROCR package \cite{Sing:2005}.

In Figure~\ref{fig:Biatech54_fully_ROC}, we see that at the species and peptide levels the AUC
computed using the ion abundance is approximately 9\% higher than that obtained using the spectral
count; the AUCs are approximately equal at the protein level. The ion abundance at the protein level,
however, yields an AUC appreciably less than that at the species and peptide levels.  By contrast,
the spectral count yields an AUC that is nearly constant across rollup levels.

Using the semi-tryptic data, the relative performance of the spectral count and ion abundance
is nearly reversed (Figure~\ref{fig:Biatech54_semi_ROC}).  Strikingly, the performance of
ion abundance at the species and peptide rollup levels is significantly lower than that
obtained using the fully-tryptic data whereas the performance at the protein level is nearly the
same. With the semi-tryptic data, as with the fully-tryptic, the performance of the spectral count
is relatively consistent across rollup levels and approximately equal to the AUCs computed using
the fully-tryptic data.

Initially, we were puzzled by these findings, particularly by the disparity of performance,
between the fully-tryptic and semi-tryptic data, of ion abundance at the species
and peptide levels of rollup.  Also confusing is that the performance of ion abundance
at the protein level of rollup in the semi-tryptic data `recovers' to that obtained with the
fully-tryptic data.

\begin{figure}[!h]
 \begin{center}
   \subfigure[]
      {\label{fig:Biatech54_strictly_semi_tryptic_obs}
       \includegraphics[width=0.9\linewidth]{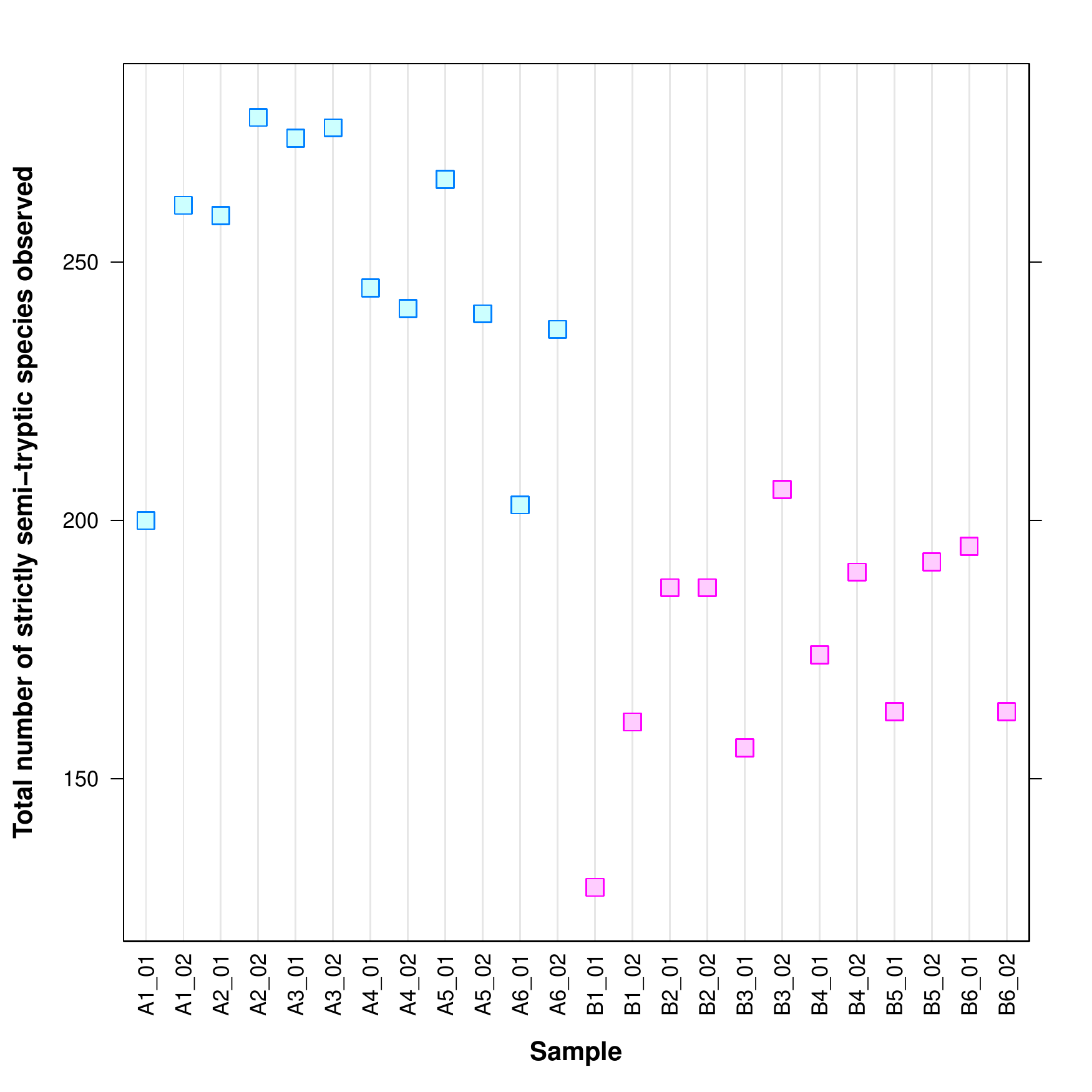}} \\
   \vspace{-0.1in}
   \subfigure[]
      {\label{fig:Biatech54_fraction_strictly_semi_tryptic_areas_signal}
       \includegraphics[width=0.9\linewidth]{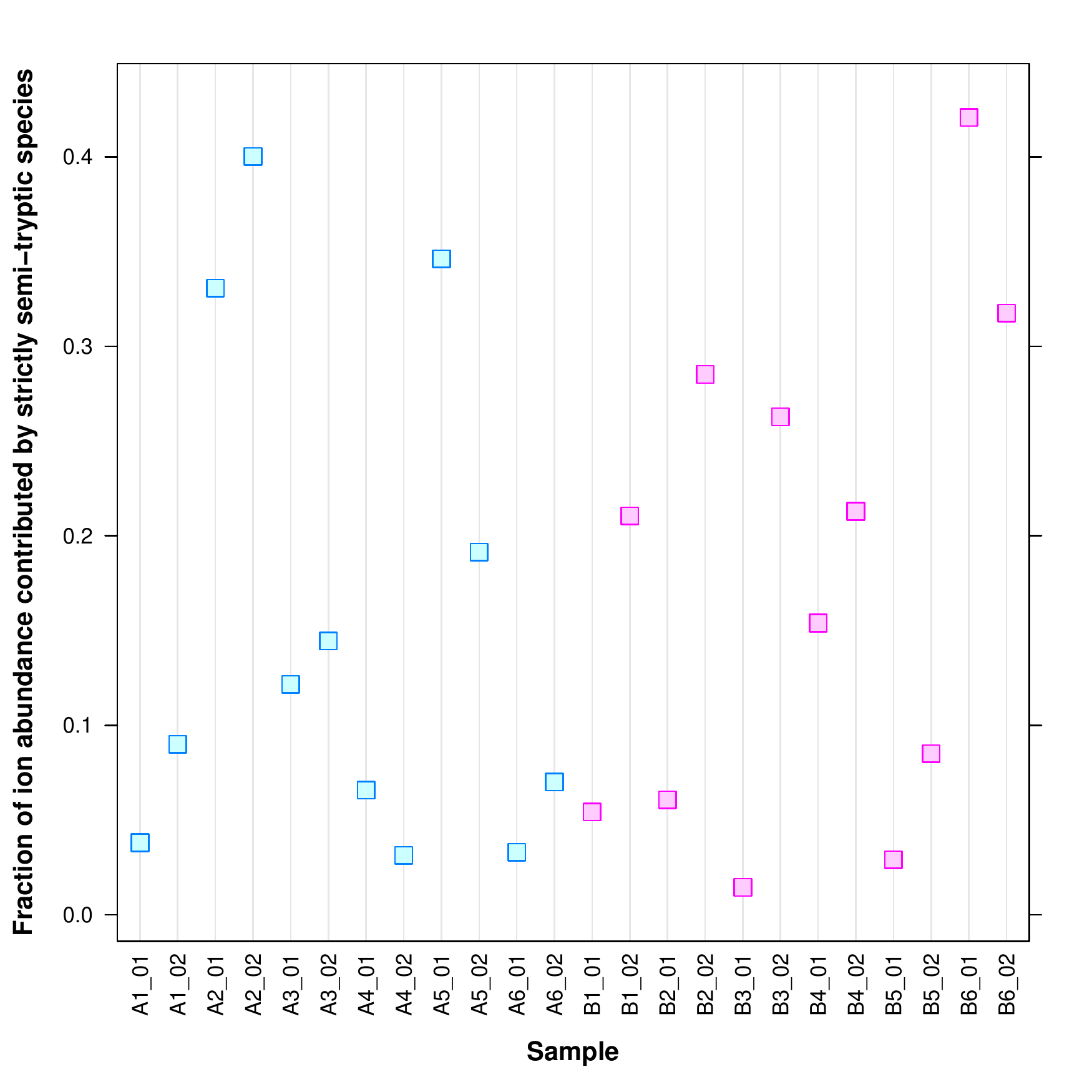}} \\
 \end{center}
 \vspace{-0.1in}
 \caption{{\bf Strictly semi-tryptic species content by experiment.} (a) Strictly semi-tryptic species are observed more
 frequently in the Mix 1 (A) than in the Mix 2 (B) samples of the BIATECH-54 data set, searched to
 permit the observation of such species. (b) The fraction of the total ion abundance contributed by
 strictly semi-tryptic species varies little between the Mix 1 and Mix 2 samples.}
 \label{fig:Biatech54_semi_tryptic}
\end{figure}

A further examination revealed that these findings reflect an actual difference between the
Mix 1 and Mix 2 samples of the BIATECH-54 set not detected using the spectral count.
Figure~\ref{fig:Biatech54_strictly_semi_tryptic_obs} shows the total number of strictly semi-tryptic
species (i.e., not fully-tryptic) detected in each of the 24 samples of the BIATECH-54 data
set.  Clearly, strictly semi-tryptic species are observed more frequently in the Mix 1 than
in the Mix 2 samples.  We believe this fact reflects a difference in the way in which the
samples themselves were prepared; a contaminating (non-trypsin) proteolytic enzyme may
have been introduced into the Mix 1 samples, or may have been more active in these than
in the Mix 2 samples.

This difference between the Mix 1 and Mix 2 samples is manifest in the ROC plots for the
semi-tryptic data because the structure of $\tau_r$ as an average of Wilcoxon statistics makes
it sensitive to the contributions made by the additional strictly semi-tryptic species observed in Mix 1.
For $\tau_s$, for example, each observation of a strictly semi-tryptic species in Mix 1 that is not
in Mix 2 makes an incremental contribution that carries equal weight to the contributions made
by species observed in both samples.

Figure~\ref{fig:Biatech54_fraction_strictly_semi_tryptic_areas_signal} shows the fraction of ion abundance
contributed by the strictly semi-tryptic species in the 24 samples.  Despite the fact that more strictly
semi-tryptic species are present in Mix 1 than in Mix 2, their relative contribution to the total ion
abundance in each sample is nearly equal across samples.  This suggests that the ion abundance
signal associated with strictly semi-tryptic species is small.  It also explains the `recovery' of performance
observed in Figure~\ref{fig:Biatech54_semi_ROC} for data analyzed at the
protein rollup level.  Rollup to the protein level `hides' the relatively small contribution made to each
protein's total ion abundance by the low-abundance strictly semi-tryptic species.

It is more difficult to provide a simple explanation for the poor performance seen in
Figure~\ref{fig:Biatech54_semi_ROC} for the intermediate, peptide level of rollup.
The performance is the result of a complex interaction of two factors: the number of species that
contribute to the quantification of each peptide; and the relative contribution made by the additional
strictly semi-tryptic species observed in the Mix 1 samples to the quantification of each peptide.

Using a semi-tryptic search strategy and the ion abundance measure, we've found that the BIATECH-54
samples Mix 1 and Mix 2 differ markedly in their content of strictly semi-tryptic species
(Figure~\ref{fig:Biatech54_strictly_semi_tryptic_obs}).  We detected this anomaly at the species and
peptide levels of rollup, but not at the protein level (Figure~\ref{fig:Biatech54_semi_ROC}), using
our Wilcoxon rank-sum-based test.  The same analysis carried out using the spectral count measure
gave no hint of this unanticipated finding.  Our analysis suggests not only that the ion abundance
is a more sensitive measure of quantification than is the spectral count, but that analysis of
LC-MS/MS data at different levels of rollup can be informative.

\subsection{CPTAC}
In a study conducted by the Clinical Proteomic Technologies for Cancer (CPTAC) consortium,
Paulovich et al.\ \cite{Paulovich:2010} introduced a reference data set measuring new
performance standards for benchmarking of LC-MS/MS platforms and data analysis methods.
These new standards are based on the yeast proteome and the UPS1 (Universal Proteomcs Standard
Set 1) collection of 48 human source or human sequence recombinant proteins \cite{UPS1:2011}.
The CPTAC reference samples and data set therefore provide a more complex and challenging
benchmark for LC-MS/MS analysis than does BIATECH-54.

\begin{table}[!h]
\caption{The composition of the CPTAC samples.}
{\begin{tabular}{lccc | c}
\toprule %
Sample & Yeast           & UPS1 (Sigma-48) \\
             & (ng/$\mu$L) & (fmol/$\mu$L)      \\
\hline
QC2      & 60   & 0       \\
A           & 60   & 0.25  \\
B           & 60   & 0.74  \\
C           & 60   & 2.2    \\
D           & 60   & 6.7    \\
E           & 60   & 20     \\
\hline
\end{tabular}}
\label{table:CPTAC design}
\end{table}

We analzyed a subset of the data collected for the CPTAC study, comparing the trypsin-digested yeast
protein lysate samples, designated QC2, with the UPS1 spike-in samples, designated A, B, C, D and E.
The composition of these samples is summarized in Table~\ref{table:CPTAC design}, which is excerpted
from Section~C of \citep[][Supplementary Information]{Paulovich:2010}.  The four laboratories
participating in the study each collected three technical replicate observations of the QC2, A, B, C, D and
E samples for a total of twelve (12) observations of each.

\begin{table}[htb] % [h(ere) t(op) b(ottom)] = order of priorities in which latex tries to place the table.
\caption[]{The number of yeast and UPS1 proteins found to be present in greater ($\Uparrow$) and lesser
($\Downarrow$) abundance in samples A-E relative to QC2 of the CPTAC data at FDR level
0.05.}
\label{table:Performance}
% \vspace{-10pt} %  this would tighten up the space between the caption and table.
% \resizebox{\textwidth}{!}{ % this resizes the table width to match the document's textwidth.
\scalebox{.875}{ % begin scalebox
\begin{tabular}{ll c  lll >{\columncolor[gray]{.85}[7pt] [2pt]}l  l} % to remove column color, replace this by {lccccc|c}
\toprule %
  &  & sample   & \multicolumn{2}{c}{\underbar{ UPS1 }}  & \multicolumn{2}{c}{\underbar{ \ yeast \ }} & total \\
  &  & (vs QC2)  & $\Uparrow$   & $\Downarrow$   & $\Uparrow$   & $\Downarrow$ &      \\
 \toprule %
 \multirow{10}{5pt}{\kern-.5em\rotatebox{90}{\Large {\bf species} }}
& \multirow{5}{0pt}{\parbox{0pt}{\bf ion  abundance} }
           & A      & 0       & 0      & 0    & 0          & 0      \\
    &     & B     & 2       & 0      & 0    & 0          & 2      \\
    &     & C     & 17     & 0      & 0    & 5          & 22    \\
    &     & D     & 35     & 0      & 0    & 12        & 47    \\
    &     & E     & 41     & 0      & 2    & 347      & 390   \\
\cmidrule(l){2-8}
& \multirow{5}{1cm}{\parbox{10pt}{\bf spectral count}  }
            & A      & 0     & 0      & 0    & 1          & 1     \\
    &      & B     & 3     & 0      & 0    & 1          & 4    \\
    &      & C     & 21   & 0      & 2    & 7          & 30    \\
    &      & D     & 35   & 0      & 2    & 4          & 41    \\
    &      & E     & 42   & 0      & 27  & 69        & 138   \\
 \toprule %
 \multirow{10}{5pt}{\kern-.5em\rotatebox{90}{\Large {\bf peptides} }}
& \multirow{5}{0pt}{\parbox{0pt}{\bf ion  abundance} }
           & A      & 0      & 0      & 0    & 0          & 0     \\
    &     & B     & 3      & 0      & 0    & 0          & 3     \\
    &     & C     & 16    & 0      & 0    & 5          & 21    \\
    &     & D     & 35    & 0      & 0    & 18        & 53    \\
    &     & E     & 41    & 0      & 2    & 403      & 446   \\
\cmidrule(l){2-8}
& \multirow{5}{1cm}{\parbox{10pt}{\bf spectral count}  }
            & A      & 0     & 0      & 0    & 1        & 1      \\
    &      & B     & 8     & 0      & 1    & 0        & 9      \\
    &      & C     & 23   & 0      & 4    & 9        & 36    \\
    &      & D     & 35   & 0      & 5    & 6        & 46    \\
    &      & E     & 42   & 0      & 20  & 71      & 133   \\
 \toprule %
  \multirow{10}{5pt}{\kern-.5em\rotatebox{90}{\Large {\bf proteins} }}
& \multirow{5}{0pt}{\parbox{0pt}{\bf ion  abundance} }
           & A      & 0      & 0      & 0    & 0         & 1      \\
    &     & B     & 3      & 0      & 0    & 0         & 3      \\
    &     & C     & 15    & 0      & 0    & 5         & 20    \\
    &     & D     & 35    & 0      & 1    & 31       & 67    \\
    &     & E     & 40    & 0      & 5    & 522     & 567   \\
\cmidrule(l){2-8}
& \multirow{5}{1cm}{\parbox{10pt}{\bf spectral count}  }
            & A      & 0     & 0      & 0    & 0          & 0     \\
    &      & B     & 5     & 0      & 0    & 1          & 6    \\
    &      & C     & 24   & 0      & 1    & 8          & 33    \\
    &      & D     & 34   & 0      & 1    & 6          & 41    \\
    &      & E     & 42   & 0      & 11   & 83        & 136   \\
 \bottomrule %

\end{tabular}
 } % end of scalebox
% }% end of resizebox
\end{table}

The MS/MS spectra were searched using X!Tandem against the same protein database used
by the CPTAC authors, which includes both target and decoy (reversed) protein
sequences. The precursor matching tolerance was set to [-10 ppm, +10 ppm].  The search was
performed to allow for potential oxidation of methionines ($\Delta m = 15.9949  \mathrm{D}$)
and carbamidomethylation of cysteines ($\Delta m = 57.0215 \mathrm{D}$).  Up to two missed
tryptic cleavages were permitted.  The search was conducted allowing for only fully-tryptic
species.

The CPTAC data was quantified by spectral count and ion abundance using Sahale executed
with the FDR threshold parameter set to 0.001.  Species not observed or quantified
in at least three (3 of 12) of the A, B, C, D
or E samples or three (3 of 12) of the QC2 samples were filtered out, as were uninformative species
in the ion abundance data for which the ion abundances were effectively zero or were missing for
all samples. The data was then rolled up.  The spectral counts in each sample were normalized
by the total spectral count, and the ion abundances by the median ion abundance.

Following these preparatory steps, we searched for proteins present in differing abundance in the
QC2 and A, B, C, D or E samples using the statistic $\tau_r$ (Equation \eqref{eq:tau_definition})
computed at the species, peptide and protein rollup levels.  Following the methodology used by Paulovich
\emph{et al}.\ \cite{Paulovich:2010} in the analysis leading to their Table~III, we treated the 12 observations of
each sample type as biological replicates. Note that because we tested on the rank computed for the QC2
samples, species, peptide or protein elements \emph{less abundant} in the A-E samples
than in QC2 have $w > 0$.  This observation will be important in the following.

\begin{figure}[!h]
 \begin{center}
   \includegraphics[width=0.9\linewidth]{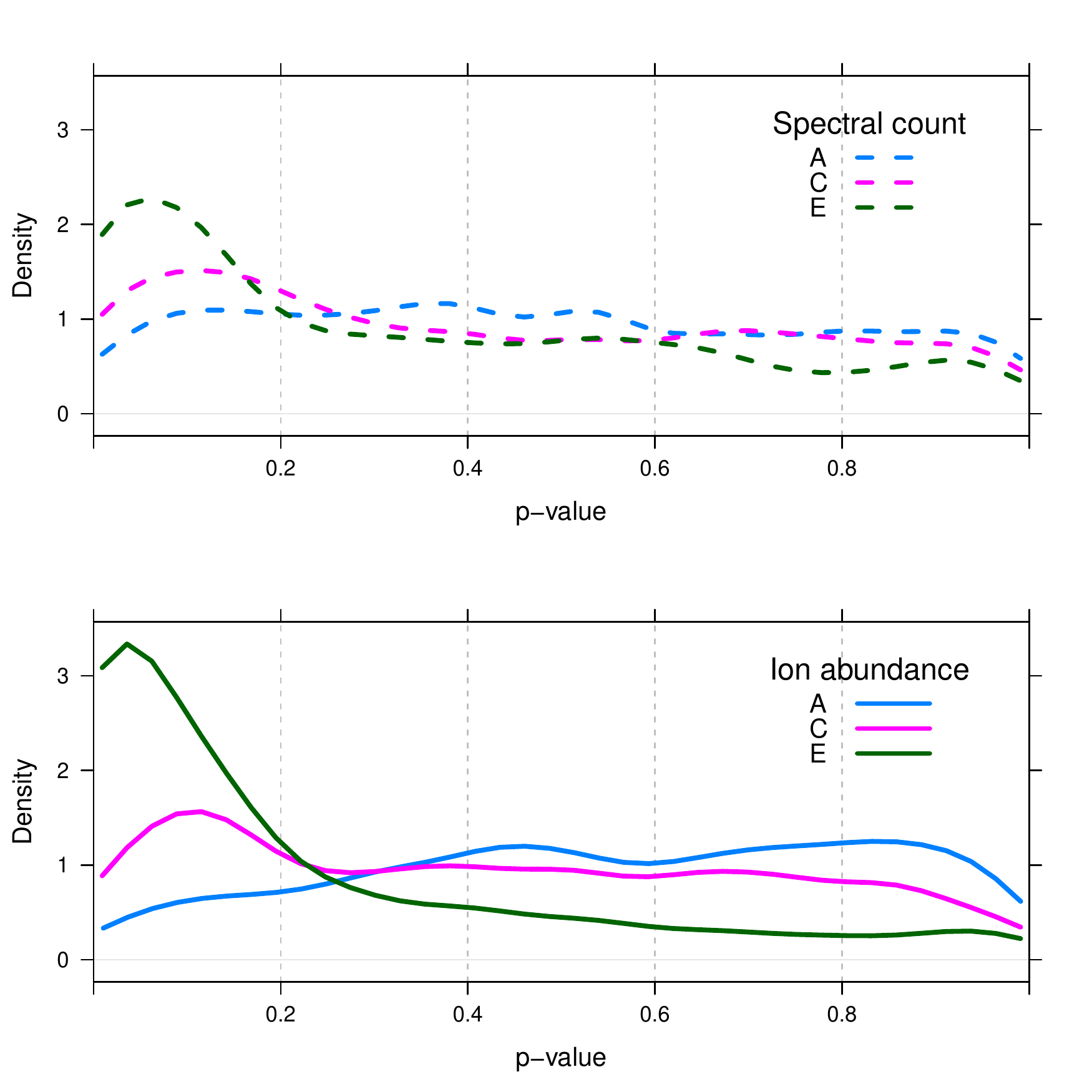}
 \end{center}
 \vspace{-0.1in}
 \caption{{\bf Stratified p-values for yeast proteins.} Density estimate of the distribution of
 $p_s$ for yeast proteins only of the A, C and E samples, tested against QC2, of the CPTAC
 data set.}
 \label{fig:CPTAC_ACE_yeast_pvalues_distribution}
\end{figure}

To correct for the effects of multiple hypothesis testing, we converted p-values to false discovery
rates using the R package \emph{qvalue} (see \cite{Storey:2003}).  The numbers
of yeast and UPS1 proteins found to be present in greater and lesser abundance in samples
A-E relative to QC2 at FDR level 0.05 are shown in Table~\ref{table:Performance}.

Table~\ref{table:Performance} shows that the performance of the spectral count is somewhat
better than ion abundance in correctly characterizing the abundance of the UPS1 proteins in
the case samples across rollup levels, particularly at the lower levels of spike-in.  However, what
stands out in Table~\ref{table:Performance} is the large number of yeast proteins found to have
different abundance in the E and QC2 samples.  The number of these apparent false positives is particularly
high in the results computed using ion abundance. We sought to understand the origin of these
'false positives', initially suspecting some error in the method by which features were quantified
by ion abundance.

Figure~\ref{fig:CPTAC_ACE_yeast_pvalues_distribution} shows the distribution of p-values
computed for the yeast proteins only of the A, C and E samples.  These p-values were computed
on the basis of $\tau_s$, that is, on the basis of the spectral counts and ion abundances
at the species level of rollup.  Figure~\ref{fig:CPTAC_ACE_yeast_pvalues_distribution} shows that
the proportion of yeast proteins determined to be significant increases as a function of increasing
UPS1 protein concentration.   This is the case for p-values computed using either the spectral
count or ion abundance measure, though the trend is more easily seen in the ion abundance results.

\begin{figure}[!h]
 \begin{center}
    \includegraphics[width=0.9\linewidth]{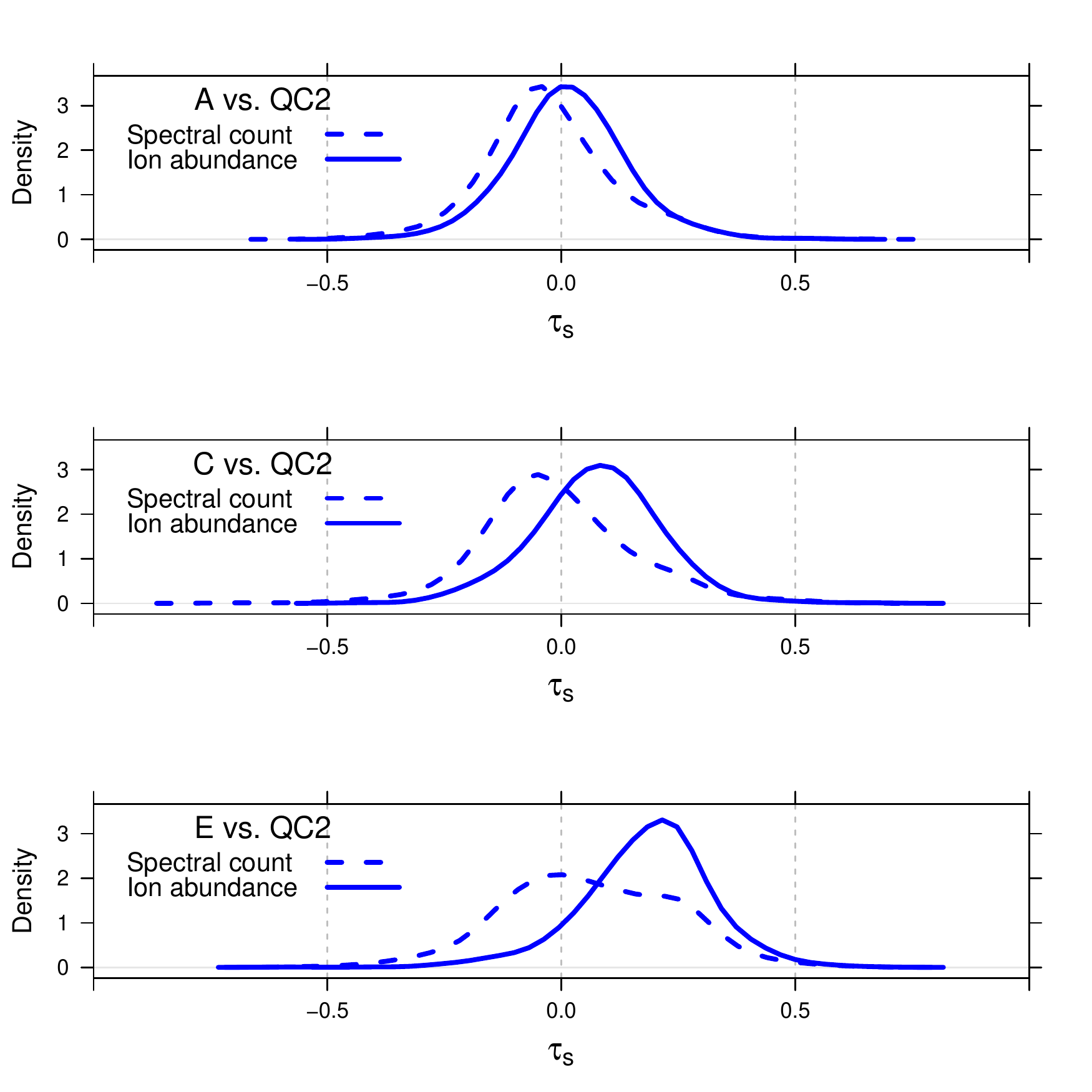}
 \end{center}
 \vspace{-0.1in}
 \caption{{\bf Stratified $\mathbf{\tau_s}$ for yeast proteins.} Density estimate of the distribution of
 $\tau_s$ for yeast proteins only of the A, C and E samples, tested against QC2, of the
 CPTAC data set.}
 \label{fig:CPTAC_yeast_tau_distribution}
\end{figure}

The results presented in Table~\ref{table:Performance} and
Figure~\ref{fig:CPTAC_ACE_yeast_pvalues_distribution} were computed on the basis of
$\tau_s$.  Figure~\ref{fig:CPTAC_yeast_tau_distribution} shows the distribution of $\tau_s$ for
the yeast proteins only in the A, C and E samples. Note that the mass of the $\tau_s$ distributions,
computed using either the spectral count or ion abundance, trend to the right as the concentration
of UPS1 proteins is increased.

The simple structure of $\tau_s$ as an average of Wilcoxons allowed us to quickly diagnose
the origin of this trend: the distribution of $w$ for yeast species must mirror that of $\tau_s$
for yeast proteins in the A, C and E samples.

\begin{figure}[!h]
 \begin{center}
   \subfigure[]
     {\label{fig:CPTAC_wilcoxon_by_QC2_signal_class}
       \includegraphics[width=0.9\linewidth]{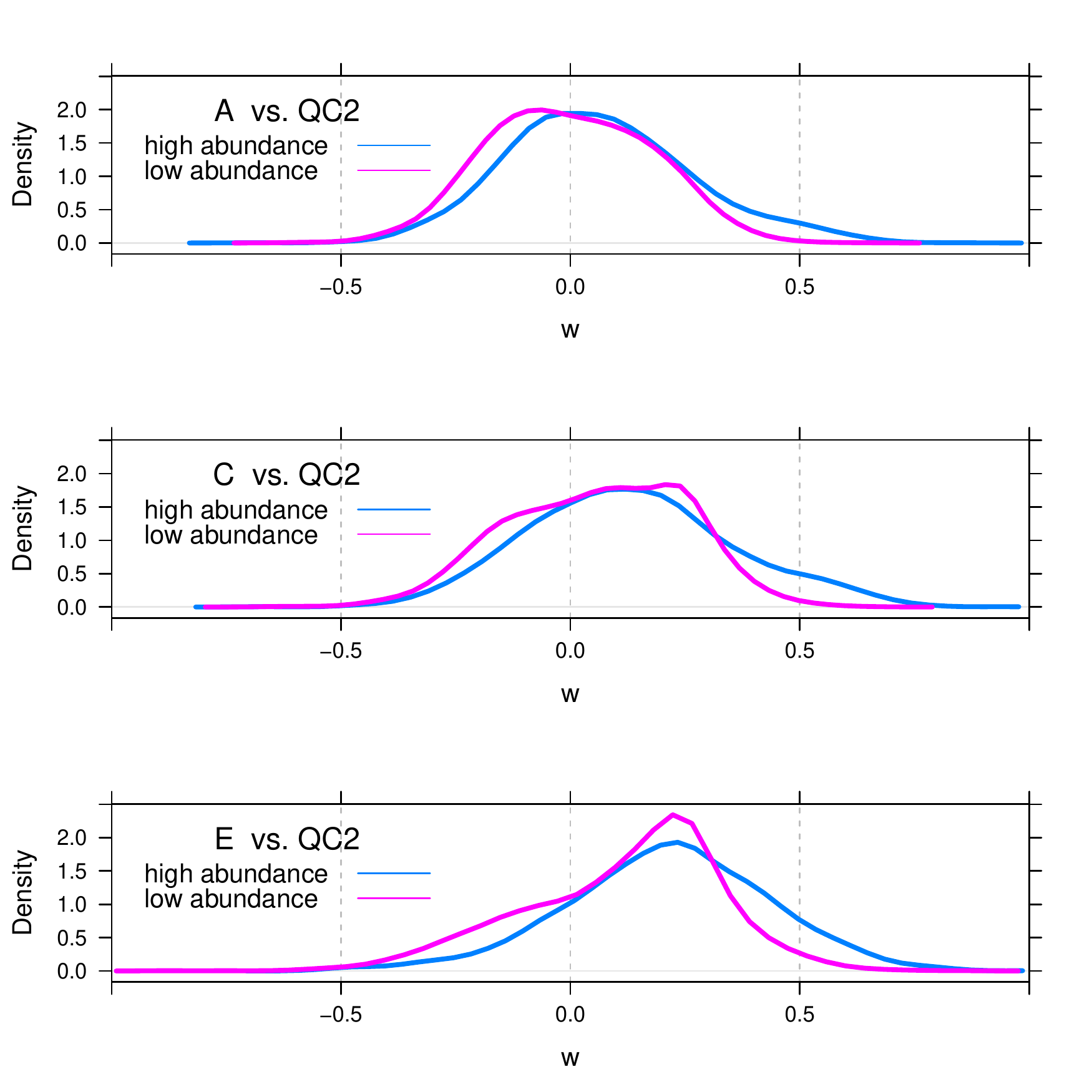}} \\
   \vspace{-0.1in}
   \subfigure[]
     {\label{fig:CPTAC_wilcoxon_by_QC2_missing_class}
      \includegraphics[width=0.9\linewidth]{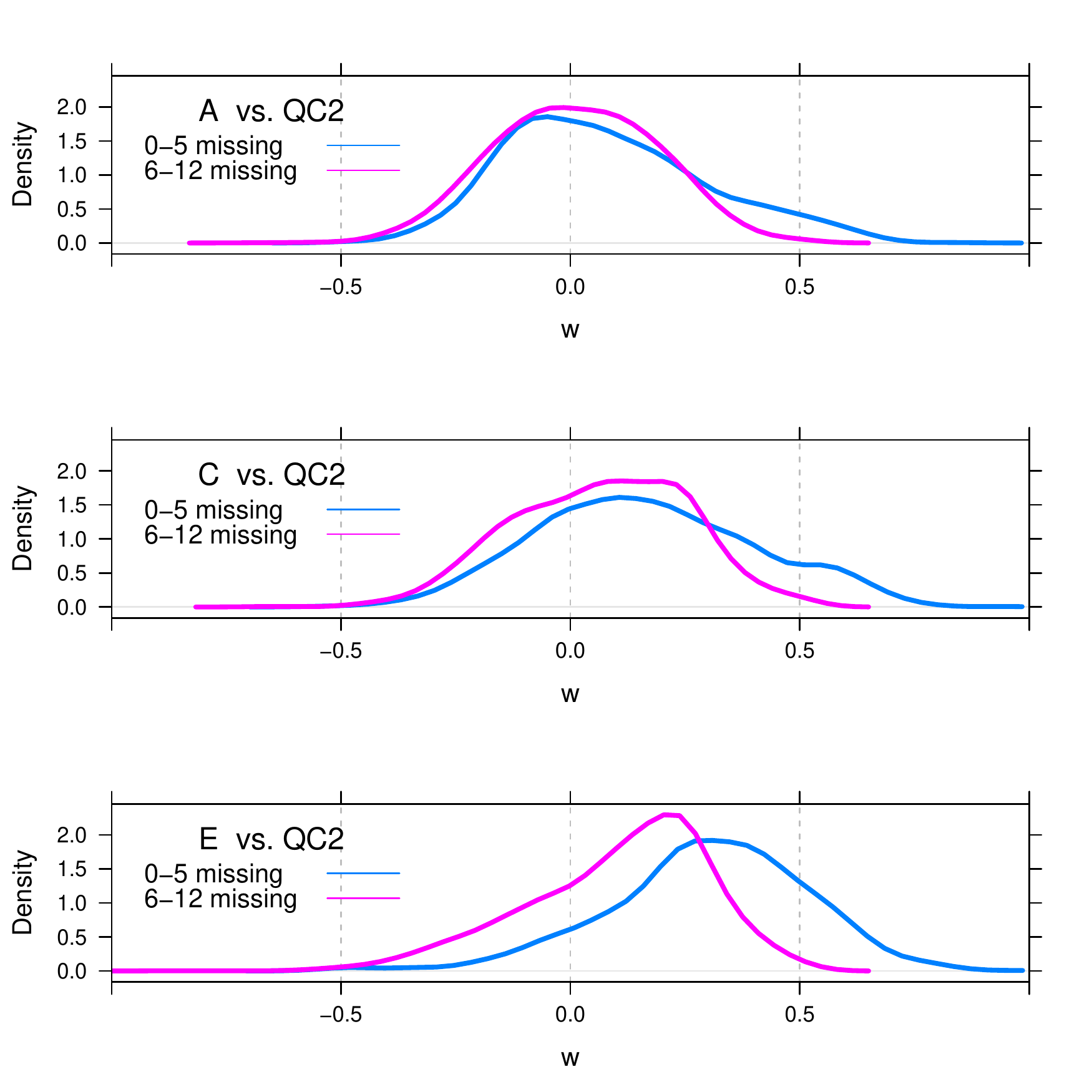}} \\
 \end{center}
 \vspace{-0.1in}
 \caption{{\bf Stratified $w$ for yeast species.} Density estimate of the distribution of the
 Wilcoxon statistic $w$ for yeast species stratified by (a) the total ion abundance (low abundance
 is $\le$ the median; high abundance is $>$ median) and (b) the total number of missing observations
 (0-5 and 6-12), both as measured in the QC2 samples.}
 \label{fig:CPTAC_wilcoxon_by_signal_and_missing_class}
\end{figure}

To determine the origin of the positive shift for $w$, we stratified $w$ by two factors, the total ion abundance
of the yeast species, and the number of missing observations of the yeast species, both
in the QC2 samples.  The positive trend of $w$ does not appear to correlate with total ion abundance
(Figure~\ref{fig:CPTAC_wilcoxon_by_QC2_signal_class}).  The distribution of $w$ for few and many
missing observations both trend positive as the UPS1 protein concentration increases, but the Wilcoxons
for yeast species with many missing observations in QC2 appear to be influenced less
(Figure~\ref{fig:CPTAC_wilcoxon_by_QC2_missing_class}). We conclude, however, that
neither factor explains the shift of $w$.

 \begin{figure}[!h]
 \begin{center}
  \includegraphics[width=1.0\linewidth]{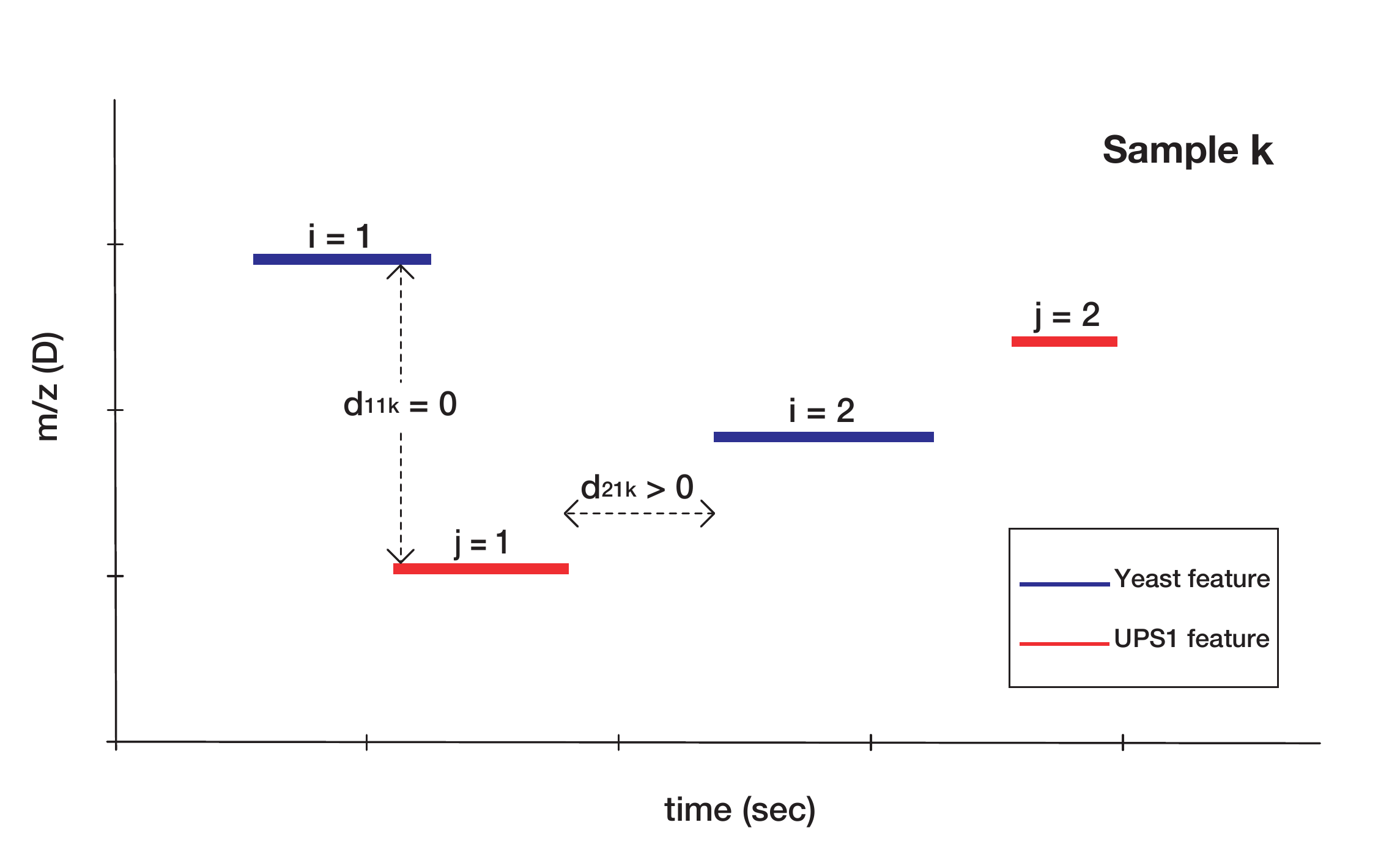}
 \end{center}
 \vspace{-0.2in}
 \caption{{\bf Schematic defining the distance between LC-MS features.} $d_{ijk}$ quantifies the
 time separation of the $i$'th yeast feature from the $j$'th UPS1 feature in sample
 $k$ (see Equation~\eqref{eq:dijk_definition}). The \textcolor{blue}{blue} and
 \textcolor{red}{red} lines represent the $2\sigma$ extent of yeast and UPS1 features,
 respectively.}
 \label{fig:CPTAC_yeast_ups1_distance_definition}
\end{figure}

The positive shift in the distribution of $w$ for the yeast species, and $\tau_s$ for the yeast proteins,
obviously correlates with the increasing concentration of the UPS1 proteins.  Consequently, we
speculated that the trend is the result of a competition for ions between yeast and UPS1 species
that elute in the same time window of the LC-MS experiments measuring the case samples, A, B, C,
D, or E. To test this, we defined an `interference distance', $d_i$, associated with each yeast feature
$i$, which measures the average (over samples) minimum time separating the $2\sigma$ extent
(see Equation~\eqref{eq:Feature function}) of a yeast feature from the $2\sigma$ extent of all UPS1 features in the
same sample. $d_i$ is defined by
\begin{equation}
 d_i = \mean_{\substack{k \in \mathrm{case \; samples}}} \left( \min_{j \in \mathrm{UPS1 \; feature}} d_{ijk} \right), \; i \in \mathrm{yeast \; features,}
 \label{eq:yeast_UPS1_distance}
\end{equation}
where $d_{ijk}$ is (roughly) the absolute value of the time separating the $i$'th yeast feature from
the $j$'th UPS1 feature in sample $k$.  If the $2\sigma$ time extent of yeast feature $i$ is denoted
$[y_L, y_R]$, and the $2\sigma$ time extent of UPS1 feature $j$ is denoted $[u_L, u_R]$, both in
sample $k$, $d_{ijk}$ is given by
\begin{equation}
 d_{ijk} =
   \begin{cases}
     y_L - u_R &\text{if $y_L > u_R$,}  \\
     u_L - y_R &\text{if $u_L > y_R$,} \\
     0                  &\text{otherwise}.
   \end{cases}
   \label{eq:dijk_definition}
\end{equation}
$d_{ijk} = 0$ when the yeast and UPS1 features overlap in time and potentially compete for ions;
$d_{ijk}$ is taken as missing if the yeast feature $i$ or UPS1 feature $j$ are missing in sample $k$.
The definition of $d_{ijk}$ is illustrated in Figure~\ref{fig:CPTAC_yeast_ups1_distance_definition}.

\begin{figure}[!h]
\begin{center}
   \includegraphics[width=0.9\linewidth]{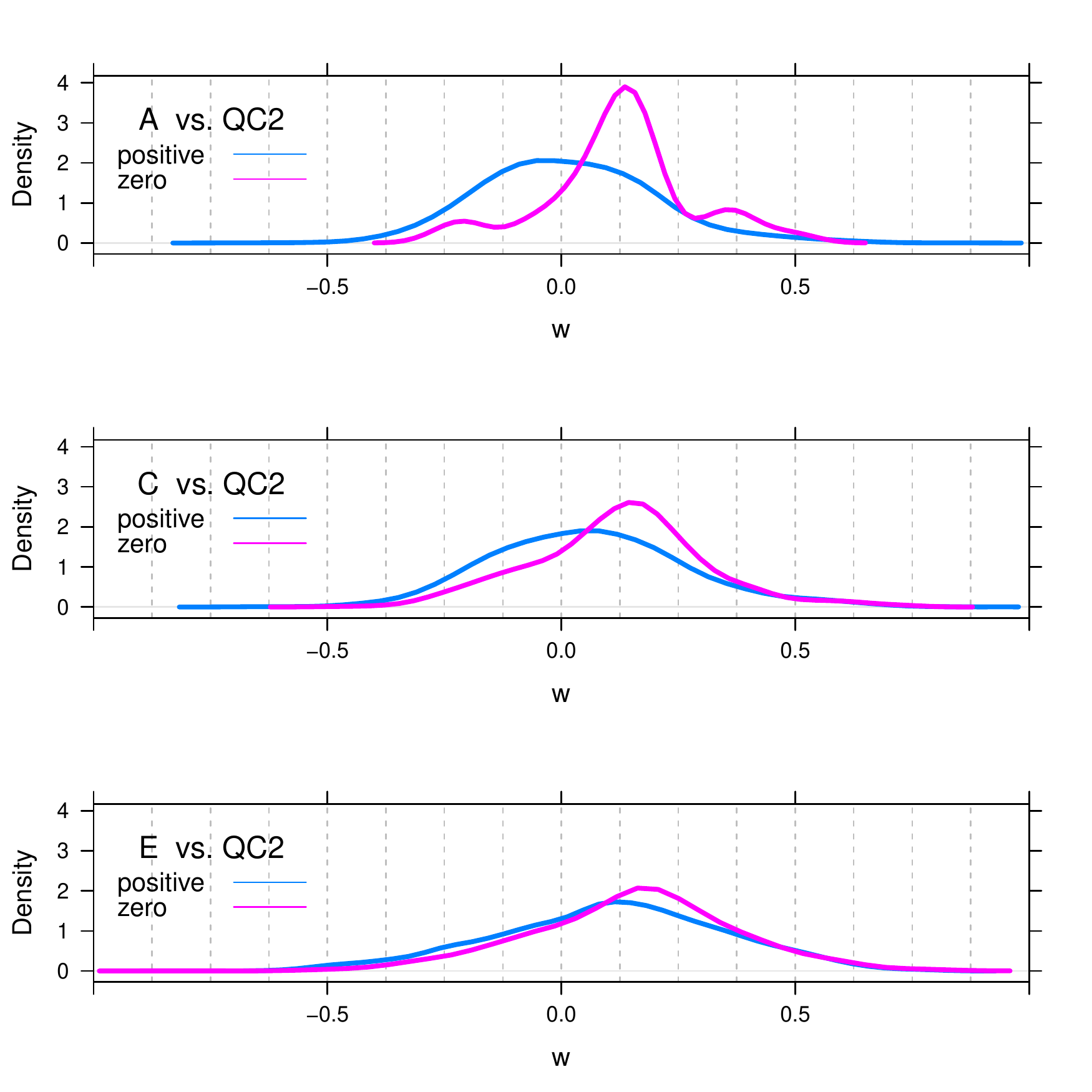}
\end{center}
\caption{{\bf Ion competition.} Density estimate of the distribution of the Wilcoxon statistic, $w$,
 for yeast species stratified by the interference distance of the yeast features from UPS1 features.}
\label{fig:CPTAC_wilcoxon_by_yeast_ups1_distance}
\end{figure}

Figure~\ref{fig:CPTAC_wilcoxon_by_yeast_ups1_distance} shows the distribution of $w$ for the yeast species in the
A, C and E samples stratified by interference distance. Table~\ref{table:interference_distance_class_populations}
shows the corresponding number of yeast species in three interference distance cohorts for samples A-E.
The `zero' cohort describes yeast species for which the interference distance $d_i = 0$; the `positive' cohort
includes yeast species with $d_i > 0$; and the `missing' cohort includes yeast species for which an interference
distance could not be computed.

Note in Figure~\ref{fig:CPTAC_wilcoxon_by_yeast_ups1_distance} that the Wilcoxons $w$ in the zero cohort
are shifted to positive values for all of the A, C and E samples.  That is, the Wilcoxon is shifted for all yeast
species whose feature overlaps in time with a UPS1 feature. This is consistent with our hypothesis that
ion competition is responsible for the positive shift in the distribution of $w$ for the yeast species and,
consequently, the positive shift in the distribution of $\tau_s$ for the yeast proteins as well.

The Wilcoxons for yeast species in the positive cohort are centered at zero in sample A and shift slowly to
more positive values as the UPS1 spike-in level increases. We hypothesized that this slow rightward shift
is a result of ion competition outside the $2\sigma$ extent we assumed in assigning yeast species to the
zero cohort.

\begin{figure}[!h]
 \begin{center}
   \includegraphics[width=0.9\linewidth]{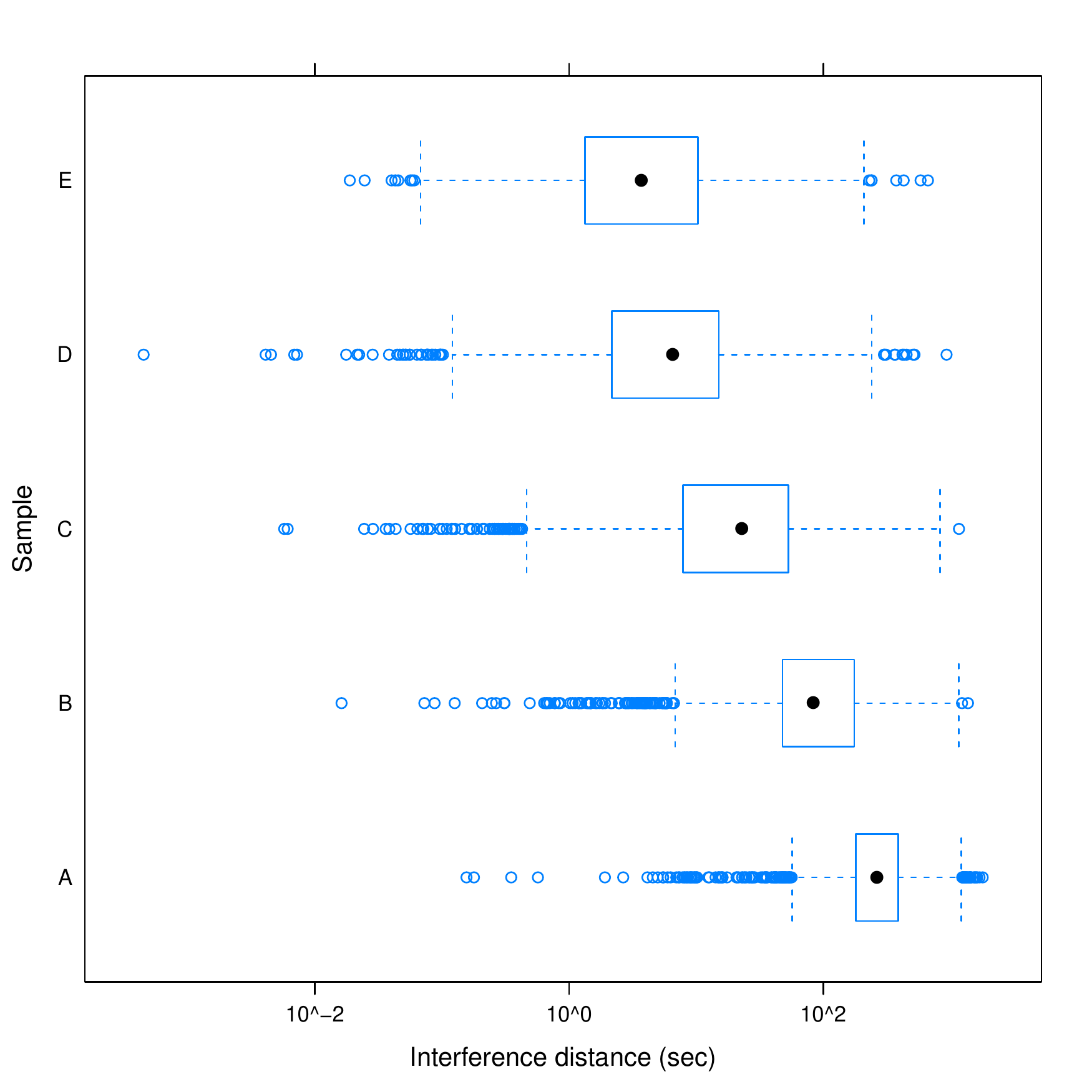}
 \end{center}
 \caption{{\bf Positive cohort interference distances.} The distribution of the interference distance
 for yeast species belonging to the positive cohort for samples A-E.}
 \label{fig:CPTAC_pos_interference_distance_dist}
\end{figure}

To test this, we plotted (Figure~\ref{fig:CPTAC_pos_interference_distance_dist}) the distribution of the interference
distance for yeast species of the positive cohort only.  Especially in sample E, the bulk of yeast features corresponding
to species in the positive cohort are in close proximity to a UPS1 feature.  This provides additional evidence that ion
competition is responsible for the rightward shift of $w$ seen in Figure~\ref{fig:CPTAC_wilcoxon_by_yeast_ups1_distance}.
Moreover, among the yeast species in the positive cohort, their Wilcoxons and interference distances
are significantly negatively correlated, based on Spearman's rank correlation coefficient ($\rho = -0.14$, and
the p-value for $H_0: \rho = 0$ is $< 10^{-5}$).

The yeast species in the missing cohort require comment.  As is evident by examining the definition of
the interference distance $d_i$ in Equation~\eqref{eq:yeast_UPS1_distance}, in most cases that $d_i$ was
not computed, the yeast species $i$ will not have been observed in the case samples, one of A-E.
This implies that $i$ was observed only in the QC2 samples.  Consequently, for yeast species in the missing
cohort, the corresponding Wilcoxons will all be positive, i.e., these species will behave very much like
the yeast species in the zero cohort with respect to their contribution to $\tau_s$.

Note that in Table~\ref{table:interference_distance_class_populations}, excepting sample D, the
number of missing yeast species increases monotonically with the increasing level of spike-in. We
interpret this to mean that as additional UPS1 protein is introduced, not only is the ion abundance of
yeast features reduced by competition with UPS1 features, but that an increasing number of yeast
features are out-competed entirely, i.e., they are not observed, or not successfully quantified for
their ion abundance.

\begin{table}[!h]
\caption[]{The number of yeast species for which the interference distance is zero, positive
or missing in the A-E samples.}
{\begin{tabular}{lcccc}
\toprule %
Sample & zero   & positive & missing & total  \\
\hline
A           & 39     & 3699 & 203 & 3941 \\
B           & 189   & 3451 & 295 & 3935 \\
C           & 682   & 2784 & 365 & 3831 \\
D           & 1907 & 1673 & 275 & 3855 \\
E           & 2683 & 690   & 437 & 3810 \\
\hline
\end{tabular}}
\label{table:interference_distance_class_populations}
\end{table}

As we saw with the BIATECH-54 data, we are able to detect characteristics of the CPTAC data set
using the ion abundance only hinted at using the spectral count. In particular, we find that the
`false positive' detections of yeast proteins $\Downarrow$ in sample E (Table~\ref{table:Performance})
are in fact not false.  They reflect a true difference in sample E, apparently
an artifact of ion competition between the yeast and UPS1 features in the LC-MS experiment.  This
competition increases as the level of spike-in of UPS1 protein increases from A to E.

\section{Discussion}
The results of an LC-MS/MS proteomics study depend on a complex set of
choices which ultimately define the `data' that is analyzed. Here we presented
a multi-faceted examination of two benchmarking studies, BIATECH-54 and CPTAC,
and showed how various such choices affect their analysis and interpretation. Most
significantly, our analysis using the ion abundance to quantify the data exposed
properties of these data sets not readily apparent from the results of a parallel analysis
using the spectral count. We also showed that the level to which data is `rolled up'
prior to analysis may significantly affect the results and that the differences so revealed can be
informative.  Additionally, we reported on how the parameters used to execute the MS/MS
search can affect the results.  To accomplish our analysis, we introduced the straightforward
statistic $\tau_r$ (Equation \eqref{eq:tau_definition}) so as to advance our goal of contrasting
transparently the influence of quantification by spectral count and ion abundance at different
levels of data rollup.

Ion abundance is more sensitive than spectral count in detecting differences
between the Mix 1 and Mix 2 samples of the BIATECH-54 set.  Using data obtained
from a search allowing only for fully-tryptic species, ion abundance performs as well or
better at all levels of rollup; the greatest sensitivity is seen at the lower levels of
rollup (Figure~\ref{fig:Biatech54_fully_ROC}).  Using data obtained by allowing for
semi-tryptic species, the ion-abundance based $\tau_r$ detects an apparent artifact
of the manner in which the BIATECH-54 samples were prepared, something not detected
when $\tau_r$ is computed using the spectral count.  This artifact was found to
arise from the existence of approximately 10-15\% more strictly semi-tryptic species in
Mix 1 than in Mix 2 (Figure~\ref{fig:Biatech54_strictly_semi_tryptic_obs}).

The different results found using the data searched for fully-tryptic and semi-tryptic species
(Figure~\ref{fig:Biatech54_fully_ROC} and Figure~\ref{fig:Biatech54_semi_ROC}) highlights
the influence of one choice made by researchers in analyzing LC-MS/MS data.  The experimental
design of the BIATECH-54 study implicitly assumes that proteins undergo full tryptic
digestion. Constraining the MS/MS search strategy to allow for only fully-tryptic species is
therefore a natural choice. Using the less constrained semi-typtic search may seem advantageous
since one may identify additional species.  However, the advantages of a less-constrained
search come not from finding additional high-quality hits, but from a better estimation of the
null, or noise, distribution that is needed to define positive hits \cite{Ding:2008}. Once the
search is complete, strictly semi-tryptic hits should be filtered out \cite{Elias:2007}.

In the CPTAC study, we detected evidence of increasing ion competition as the
amount of UPS1 spike-in protein increases. This is weakly detectable when
$\tau_r$ is computed using the spectral count but is obvious when the ion
abundance is used; see Table~\ref{table:Performance} and the companion
perspectives in Figures~\ref{fig:CPTAC_ACE_yeast_pvalues_distribution} and
\ref{fig:CPTAC_yeast_tau_distribution}.  The sample-dependent trends seen in
these figures reveals that the abundance of yeast proteins decreases as a function
of increasing UPS1 spike-in.  Conceivably, this was an artifact of a quantification
method that treated high/low abundance or missing features differently.  However,
by stratifying the Wilcoxon $w$ on low-versus-high total ion abundance and on
few-versus-many missing observations of yeast species in QC2
(Figure~\ref{fig:CPTAC_wilcoxon_by_signal_and_missing_class}), we diagnosed
that this is not the case.  The existence of ion competition in the CPTAC data was confirmed
by stratifying the Wilcoxon statistics for yeast species on the basis of their interference
distance (see Figures~\ref{fig:CPTAC_wilcoxon_by_yeast_ups1_distance} and
\ref{fig:CPTAC_pos_interference_distance_dist}).

Our findings for the CPTAC data raises the question as to whether the spike-in
experimental design is appropriate for the construction of a benchmark case-control study.
The very introduction of a spike-in appears to bias the data by ion competition.  Possibly, the bias we
detect is a consequence of `too much' UPS1 protein having been introduced, in the
E sample in particular.  However, a simple calculation confirms that the average mass
of yeast protein per $\mu$l of E sample exceeds that for an average UPS1 protein by
approximately 20\% ($1.1\times 10^{-11}$ grams per UPS1 protein versus
$1.33\times 10^{-11}$ grams per yeast protein, assuming $\sim$4,500 expressed yeast
proteins \cite{Paulovich:2010}).  This suggests that one must be cautious in interpreting
the results obtained from a label-free LC-MS/MS experiment, as an over-expressed collection
of proteins may interfere with the ion signal measured for other classes.

Our findings differ somewhat from those of Zybailov et al.~\cite{Zybailov:2005} who conclude
that the spectral count is more reliable than ion abundance (as summarized by the RelEx
method of MacCoss et al.~\cite{MacCoss:2003}). Similarly, Old et al.~\cite{Old:2005} conclude
that the spectral count is more sensitive than ion abundance (as summarized by Serac) in
detecting differentially expressed proteins. On the other hand, that study also observes that
the ion abundance yields more accurate estimates of protein ratios than does the spectral count
and so no definitive conclusion was drawn. We note that Old et al. use one statistic for spectral
count data and another statistic for ion abundance data and so it is difficult to compare their
results to ours. Indeed, comparison between spectral count and ion abundance is complicated by the
myriad ways in which MS and MS/MS data are quantified and summarized statistically.  This
is an important point: neither of the terms ``spectral count" nor ``ion abundance" refers to a
well-defined quantification method but rather to a general approach used to define quantities
that enter into the statistical analysis.  Not only are there many ways to define these quantities
but there are also many ways to define the statistics that ultimately summarize protein comparisons,
as seen in our discussion of peptide ``rollup".

In our analyses, we have attempted to ensure that any differences observed when using
spectral count versus ion abundance for quantification are not due to non-comparable
aspects of the analysis. Our use of the Wilcoxon in defining $\tau_r$ for both measures of
quantification at each level of rollup allows for a statistically even-handed comparison.
Additionally, the MS/MS-directed approach we employed to quantify ion abundance ensures
the fairness of the comparison we make between the spectral count and ion abundance as this
approach quantifies the {\em same set of CIDs} in each case.

We note in passing that we investigated a variant of $\tau_r$ computed using the
$t$-statistic.  We found the performance of this variant is less attractive than that
based on the Wilcoxon.  This is a consequence of the small sample size of the case studies
and the parametric assumptions underlying the use of the $t$-statistic.

Finally, there are a variety of reasons to favor ion abundance for quantification. The statistical
models for protein rollup by Clough et al.~\cite{Clough:2009}, for example, are implicitly based
on ion abundances. Also, as noted by Podwojski et al.~\cite{Podwojski:2010} and Lundgren et
al.~\cite{Lundgren:2010}, spectral counts may be dominated by a few proteins having a large
number of counts, and the spectral count breaks down as a statistical quantity when very few
counts are observed.  Although the estimated ion abundance of an identified species is subject
to low signal and the stochastic nature of the CID sampling, it has the potential to more robustly
quantify seldom-seen species.

\section*{Acknowledgements}
The authors thank Jason Hogan and Matthew Fitzgibbon for their assistance
in preparing the BIATECH-54 and CPTAC data for analysis.

\paragraph*{\textit{Funding}:} This work was supported by NIH grants R21 RR025787 (TR,TM),
R01 CA126205 (TR,TM,PW), R01 GM082802 (PW,TR) and U01 CA086368 (TR).

\bibliographystyle{natbib}
\bibliography{Milac_SII_arXiv}

\end{document}